\documentclass[a4paper,12pt]{article}
\usepackage{slashed}
\usepackage{amsfonts}
\usepackage{multirow}
\usepackage{mathrsfs}
\usepackage{graphicx}
\usepackage{amsmath}
\usepackage{amssymb}
\usepackage{bm}
\usepackage{bbm}
\usepackage{color}
\usepackage[numbers, square, comma, sort&compress]{natbib}
\usepackage{hyperref}
\usepackage{geometry}
\usepackage[utf8]{inputenc}
\newcommand{\nn}{\nonumber}
\newcommand{\beq}{\begin{equation}}
\newcommand{\eeq}{\end{equation}}
\newcommand{\bqa}{\begin{eqnarray}}
\newcommand{\eqa}{\end{eqnarray}}

\newcommand{\GG}{\ensuremath \text{F}}
\newcommand{\M}{\ensuremath \text{M}}
\newcommand{\N}{\ensuremath \text{N}}
\newcommand{\BB}{\ensuremath \text{B}}

\allowdisplaybreaks[4]

\begin{document}

\begin{titlepage}
\vskip 25mm
\begin{center}
\Large\bf{\large\bf Analytic two-loop master integrals for $tW$ production at hadron colliders: I}
\end{center}
\vskip 8mm
\begin{center}
{\bf Long-Bin Chen$^{1}$, Jian Wang$^2$}\\
\vspace{10mm}
\textit{$^1$School of physics and materials science, Guangzhou University, Guangzhou 510006, China}\\
\textit{$^2$School of Physics, Shandong University, Jinan, Shandong 250100, China} \\ 
\vspace{5mm}

\end{center}
\vspace{10mm}

\begin{abstract}
We present the analytic calculation of two-loop master integrals that are relevant for $tW$ production at hadron colliders.
We focus on the integral families with only one massive propagator.
After choosing a canonical basis, the differential equations for the master integrals can be transformed into the $d$\,ln form.
The boundaries  are determined by simple direct integrations or
regularity conditions at kinematic points without physical singularities.
The analytical results in this work are expressed in terms of multiple polylogarithms,
and have been checked with numerical computations.

\end{abstract}

\end{titlepage}

\section{Introduction}

\begin{figure}[ht]
\begin{center}
\includegraphics[scale=0.4]{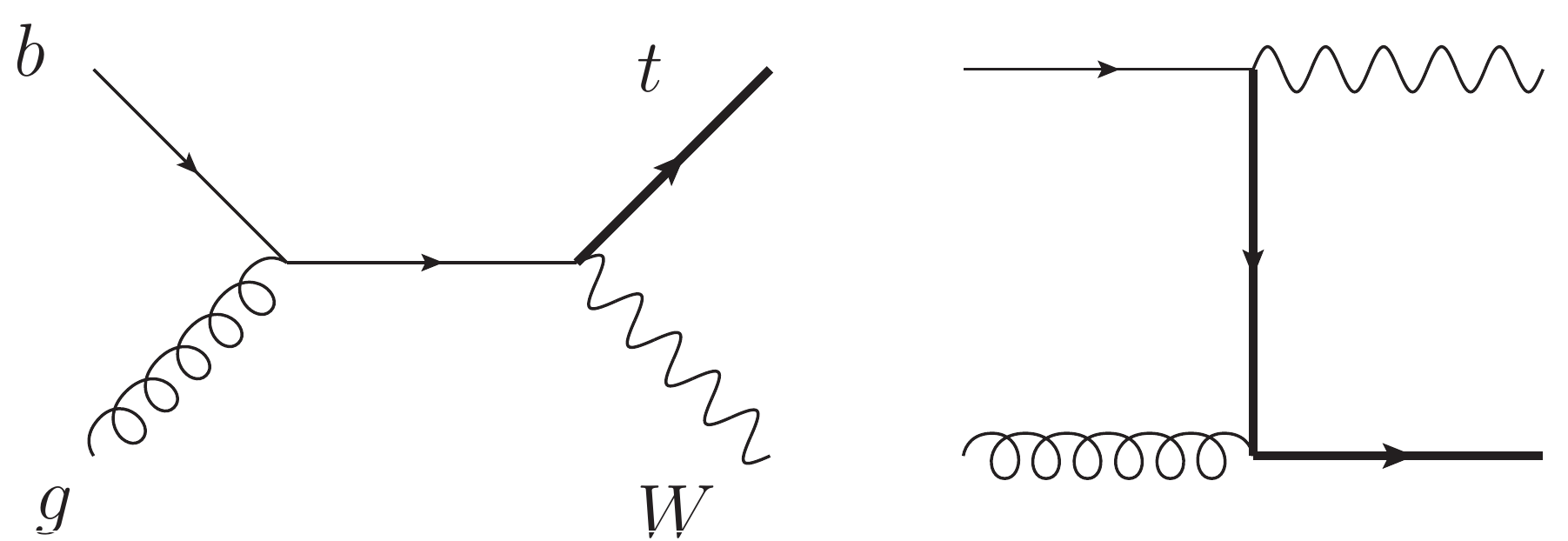}
\caption{Leading order Feynman diagrams for $gb\to tW$.}
\label{fig:LO}
\end{center}
\end{figure}

As the heaviest fundamental particle in the standard model (SM),
the top quark has played a special role in testing the structure of the SM.
It is also expected that the top quark has a close relation to new physics
because its mass is around the  scale of electroweak symmetry breaking.
Precise measurement of its properties is an important task for experiments at the 
large hadron collider (LHC).
The single top quark production can be used to detect the electroweak coupling
of top quarks, especially to determine the  Cabibbo-Kobayashi-Maskawa matrix element $V_{tb}$.
Among the three channels, the $tW$ associated production, of which the leading order Feynman diagrams are
shown in Fig.\ref{fig:LO},
 has the second largest cross section at the LHC,
making it experimentally measurable \cite{Aad:2012xca,Aad:2015eto,Aaboud:2016lpj,Aaboud:2017qyi,Chatrchyan:2012zca,Chatrchyan:2014tua,Sirunyan:2018lcp}.

In order to compare with the experimental results,
precision theoretical predictions are indispensable.
The fixed-order corrections have been computed only up to next-to-leading order in QCD
 for both the stable $tW$ final state \cite{Giele:1995kr,Zhu:2001hw,Cao:2008af, Kant:2014oha}
 or the process with their decays \cite{Campbell:2005bb}.
The parton shower and soft gluon resummation effects have been investigated in \cite{Frixione:2008yi,Re:2010bp,Jezo:2016ujg} and \cite{Li:2019dhg}, respectively.
Expanding the all-order formula of the threshold resummation to fixed orders,
the approximate next-to-next-to-next-to-leading order total cross section has been obtained  \cite{Kidonakis:2006bu,Kidonakis:2010ux,Kidonakis:2016sjf,Kidonakis:2021vob}.

In the real corrections for $tW^-$ production, there is a contribution from the 
$gg(q\bar{q})\to tW^-{\bar b}$ channel, which can interfere with the top quark pair production $gg(q\bar{q})\to t\bar{t}$ followed by the decay $\bar{t}\to W^-{\bar b}$. 
These resonance effects make the higher order correction so large that
the perturbative expansion is no longer valid.
There are several methods that have been proposed in the literature to deal with this problem. 
One can simply remove the Feynman diagrams containing two top quark resonances
if the gauge dependence is negligible \cite{Frixione:2008yi}. 
In a gauge invariant way, one could subtract the contribution of the $t\bar{t}$ on-shell production
and decay from the total $tW(b)$ cross section either globally \cite{Tait:1999cf,Zhu:2001hw} 
 or locally \cite{Frixione:2008yi,Re:2010bp}. 
 The interference can also be suppressed just by choosing special cuts on the final-state particles \cite{Belyaev:1998dn,Belyaev:2000me,Campbell:2005bb,White:2009yt} so that there is a clear definition of the $tW$ production channel.
 See \cite{Demartin:2016axk} for a review of these methods and implementation in MadGraph5\_aMC@NLO.

So far, the exact next-to-next-to-leading order QCD corrections are still unavailable, though
the next-to-next-to-leading order N-jettiness soft function of this process,
one of the ingredients for a full next-to-next-to-leading order differential calculation using a slicing method,
has been calculated in~\cite{Li:2016tvb,Li:2018tsq}.
The main bottleneck is the two-loop virtual correction, which involves multiple scales.
It is the purpose of this paper to start the first step toward tackling this problem.

The last a few decades have seen impressive progress of understanding the
structure underlying the scattering amplitude and of the calculation of multi-loop Feynman integrals.
For a specific process at a collider, the corresponding Feynman integrals can be categorized
into different families according to their propagator configurations.
And then the integrals in each family can be reduced to a small set of basis integrals,
which are called master integrals, by making use of the algebraic relations among them,
such as the identities generated via Integration by Parts (IBP) \cite{Chetyrkin:1981qh}.
The number of master integrals has proven to be finite \cite{Smirnov:2010hn}.
This IBP reduction procedure has been implemented in public computer programs,
such as {\tt AIR} \cite{Anastasiou:2004vj},  {\tt Reduze} \cite{vonManteuffel:2012np},
{\tt LiteRed} \cite{Lee:2012cn},
{\tt FIRE} \cite{Smirnov:2019qkx}, {\tt Kira} \cite{Klappert:2020nbg},
based on the Laporta algorithm \cite{Laporta:2001dd}.
As a consequence, the main task is to evaluate the master integrals either
analytically or numerically; see recent reviews  \cite{Heinrich:2020ybq,Blumlein:2021pgo}.
For multi-loop integrals with multiple scales, it turns out that the differential equation
is an efficient analytic method \cite{Kotikov:1990kg,Kotikov:1991pm} since
it avoids the direct loop integration, which is rather complicated in some cases,
by transforming the problem to finding a solution of a set of partial differential equations.
This method has become widely used in a lot of multi-loop calculations
after the observation that the differential equations
can significantly simplify after choosing a canonical basis \cite{Henn:2013pwa}.

The rest of this paper is organized as follows.
 In section \ref{sec:basis}, we present the canonical basis and the corresponding differential equations. 
 We discuss the determination of boundary conditions 
 and present the analytical results in section \ref{sec:result}. 
 The conclusion is given in section \ref{sec:conclusion}.

\section{The canonical basis and differential equations}
\label{sec:basis}

\begin{figure}[ht]
\begin{center}
\includegraphics[scale=0.5]{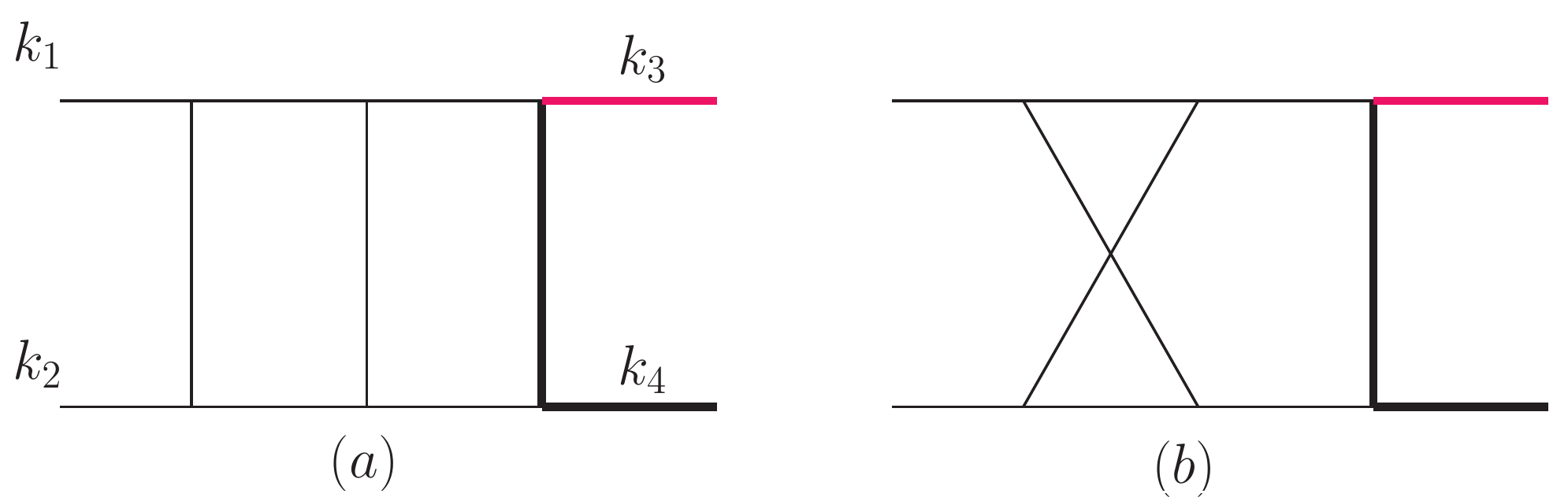}
\caption{The planar (left) and non-planar (right) diagrams of the two-loop master integrals for $gb\to Wt$ with one massive propagator.
The massive external momenta are defined by $k_3^2=m_W^2, k_4^2=m_t^2$
and we consider that $k_1,k_2$ are ingoing while $k_3,k_4$ are outgoing.}
\label{fig:massive1}
\end{center}
\end{figure}

The process of $g(k_1)b(k_2)\to W(k_3)t(k_4)$ contains two massive final states with different masses.
For the external particles, there are on-shell conditions $k_1^2=0,k_2^2=0,k_3^2=m_W^2$ and $k_4^2=(k_1+k_2-k_3)^2=m_t^2$.
The Mandelstam variables are defined as
\beq
s=(k_1+k_2)^2\,, \qquad t=(k_1-k_3)^2\,, \qquad u=(k_2-k_3)^2 \,
\eeq
with $s+t+u=m_W^2+m_t^2$.
For later convenience, we define dimensionless variables $y$ and $z$ as
 \beq
 t=y\, m_t^2,\, \quad m_W=z\, m_t\,.
\eeq

It is usually believed that the more massive propagators a diagram involves, the more complicated
the result is.
The two-loop virtual corrections can have up to four massive propagators.
Therefore, it is natural to divide the calculation to different parts according to the number of the massive propagators.
In this paper, we focus firstly on the diagrams with a single massive propagator.
Fig.\ref{fig:massive1} shows two such  diagrams with a double box topology,
one being planar while the other non-planar.
We discuss only the planar diagram in the main text, leaving the non-planar diagram to the appendix.
The amplitude of the planar diagram has been reduced to ten form factors in \cite{Basat:2021xnn}.

We define the planar integral family, including the master integral shown in Fig.\ref{fig:massive1}$(a)$,  in the form of
\bqa
I_{n_1,n_2,\ldots,n_{9}}=\int{\mathcal D}^D q_1~{\mathcal D}^D q_2\frac{1}{D_1^{n_1}~D_2^{n_2}~D_3^{n_3}~D_4^{n_4}~D_5^{n_5}~D_6^{n_6}~D_7^{n_7}D_8^{n_8}~D_9^{n_9}},
\label{def}
\eqa
with
\beq
{\mathcal D}^D q_i = \frac{\left(m_t^2 \right)^\epsilon}{i \pi^{D/2}e^{-\epsilon\,\gamma_E}}  d^D q_i \ ,\quad D=4-2\epsilon \,.
\eeq
The nine denominators are given by
\bqa
D_1&=&q_1^2,\quad D_2=q_2^2,\quad D_3=(q_1-k_1)^2,\nn \\
D_4 & = & (q_1+k_2)^2,\quad D_5=(q_1+q_2-k_1)^2,\nonumber\\
D_6&=&(q_2-k_1-k_2)^2,\quad D_7=(q_2-k_3)^2-m_t^2,\nonumber\\
D_8&=&(q_1+k_1+k_2-k_3)^2-m_t^2,\quad D_9=(q_2-k_1)^2.\nonumber
\label{int1}
\eqa
Because of momentum conservation, we do not need $k_4$ in the denominators.
The first seven  denominators can be read directly from  Fig.\ref{fig:massive1}$(a)$.
The last two are added to form a complete basis for all Lorentz scalars that can be constructed from two loop momenta and three independent external momenta.
The  denominators $D_8,D_9$ appear only with non-negative powers.
They take a form that is vanishing when the loop momentum, $q_1$ or $q_2$,
becomes soft, and therefore they are less divergent.
Besides, the choice of $D_9$ can be justified following the method in \cite{DiVita:2019lpl}.
If we put the four massless propagators containing $q_1$ on-shell,
then we get a Jacobian
\begin{align}
J=\frac{1}{(k_1+k_2)^2 (q_2-k_1)^2}.
\end{align}
From the one-loop calculation, we know the remaining three uncut propagators containing $q_2$
 already form a MI in $\epsilon$-form (up to a factor depending on the external momenta).
Therefore, a  $D_9$ in the numerator would just cancel the hidden $q_2$ propagator in the Jacobian.

\begin{figure}[ht]
\begin{center}
\includegraphics[scale=0.6]{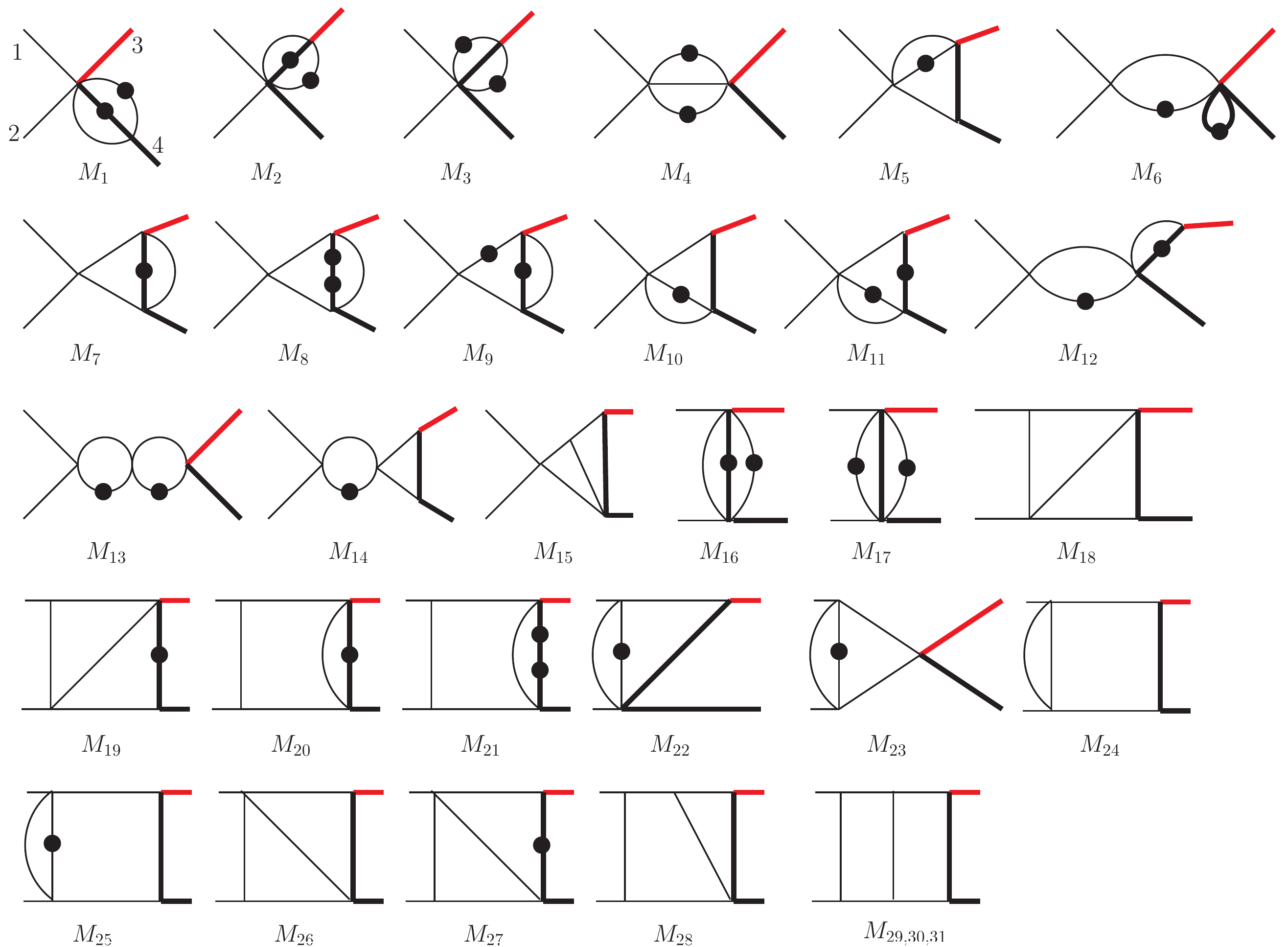}
\caption{Master integrals in the planar family.
The thin  lines  stand for massless particles while the thick lines for massive ones.
The red line in the final state denotes $W$.
Each block dot indicates one additional power of the corresponding propagator.
Numerators are not shown explicitly in the diagram and could be found in the text. }
\label{fig:planar-A}
\end{center}
\end{figure}

Making use of the {\tt FIRE} package, we find that the  integrals in the planar family
can be reduced to  a basis of 31 MIs after considering the symmetries between integrals.
We first select the MIs in such a form
 that the differential equations have coefficients linear in $\epsilon$.
These MIs  are given by
\begin{align*}
\M_{1}&=\epsilon^2 \, I_{0, 0, 0, 1, 2, 0, 2, 0, 0}\,,  &
\M_{2}&=\epsilon^2 \, I_{0, 0, 1, 0, 2, 0, 2, 0, 0}\,,  &
\M_{3}&=\epsilon^2 \, I_{0, 0, 2, 0, 2, 0, 1, 0, 0}\,,  \\
\M_{4}&=\epsilon^2 \, I_{0, 0, 1, 0, 2, 2, 0, 0, 0}\,,  &
\M_{5}&=\epsilon^3 \, I_{0, 0, 1, 0, 2, 1, 1, 0, 0}\,,  &
\M_{6}&=\epsilon^2 \, I_{0, 0, 1, 2, 0, 0, 2, 0, 0}\,,  \\
\M_{7}&=\epsilon^3 \, I_{0, 0, 1, 1, 1, 0, 2, 0, 0}\,,  &
\M_{8}&=\epsilon^2 \, I_{0, 0, 1, 1, 1, 0, 3, 0, 0}\,,  &
\M_{9}&=\epsilon^2 \, I_{0, 0, 2, 1, 1, 0, 2, 0, 0}\,,  \\
\M_{10}&=\epsilon^3 \, I_{0, 1, 0, 1, 2, 0, 1, 0, 0}\,,  &
\M_{11}&=\epsilon^2 \, I_{0, 1, 0, 1, 2, 0, 2, 0, 0}\,,  &
\M_{12}&=\epsilon^2 \, I_{0, 1, 1, 2, 0, 0, 2, 0, 0}\,,  \\
\M_{13}&=\epsilon^2 \, I_{0, 1, 1, 2, 0, 2, 0, 0, 0}\,,  &
\M_{14}&=\epsilon^3 \, I_{0, 1, 1, 2, 0, 1, 1, 0, 0}\,,  &
\M_{15}&=\epsilon^4 \, I_{0, 1, 1, 1, 1, 0, 1, 0, 0}\,,  \\
\M_{16}&=\epsilon^2 \, I_{1, 0, 0, 0, 2, 0, 2, 0, 0}\,,  &
\M_{17}&=\epsilon^2 \, I_{2, 0, 0, 0, 2, 0, 1, 0, 0}\,,  &
\M_{18}&=\epsilon^4\, I_{1, 0, 1, 0, 1, 1, 1, 0, 0}\,,  \\
\M_{19}&=\epsilon^3\, I_{1, 0, 1, 0, 1, 1, 2, 0, 0}\,,  &
\M_{20}&=\epsilon^3 \, I_{1, 0, 1, 1, 1, 0, 2, 0, 0}\,, &
\M_{21}&=\epsilon^2 \, I_{1, 0, 1, 1, 1, 0, 3, 0, 0}\, , \\
\M_{22}&=\epsilon^3\, I_{1, 1, 0, 0, 2, 0, 1, 0, 0}\,,  &
\M_{23}&=\epsilon^3 \, I_{1, 1, 0, 0, 2, 1, 0, 0, 0}\,, &
\M_{24}&=\epsilon^3(1-2\epsilon) \, I_ {1, 1, 0, 0, 1, 1, 1, 0, 0}\, , \\
\M_{25}&=\epsilon^3\, I_{1, 1, 0, 0, 2, 1, 1, 0, 0}\,,  &
\M_{26}&=\epsilon^4 \, I_{1, 1, 0, 1, 1, 0, 1, 0, 0}\,, &
\M_{27}&=\epsilon^3 \, I_{1, 1, 0, 1, 1, 0, 2, 0, 0}\, , \\
\M_{28}&=\epsilon^4\, I_{1, 1, 1, 1, 1, 0, 1, 0, 0}\,,  &
\M_{29}&=\epsilon^4 \, I_{1, 1, 1, 1, 1, 1, 1, 0, 0}\,, &
\M_{30}&=\epsilon^4 \, I_{1, 1, 1, 1, 1, 1, 1, 0, -1}\, , \\
\M_{31}&=\epsilon^4\, I_{1, 1, 1, 1, 1, 1, 1, -1, 0}\, .
\stepcounter{equation}\tag{\theequation}
\label{def:LBasisT1}
\end{align*}
The corresponding topology diagrams are displayed in Fig.\ref{fig:planar-A}.

Then we transform the MIs to a canonical basis using a method similar to that described in \cite{Argeri:2014qva},
starting from the lower sectors (with fewer propagators) to higher sectors (with more propagators).
The main logic is to treat the $\epsilon$ parts in the differential equations as perturbations.
After solving the differential equation in four dimensions, i.e., omitting the perturbations,
we get the dominant part of the MIs.
Then the full solution can be obtained by using the method of variation of constants.
The coefficient functions varied from the constants
obey the canonical form of differential equations.
For the integrals in the same sector, we have chosen a basis such that the
differential equations vanishing in four dimensional spacetime.
For example,  $F_2$ and $F_3$ belong to the same sector.
They satisfy differential equations
\bqa
\frac{\text{d} \M_{2} }{\text{d} z}&=&-\frac{2(1+\epsilon)}{z}\M_2-\frac{2\epsilon}{z}\M_3,\nonumber\\
\frac{\text{d} \M_{3} }{\text{d} z}&=&\left(\frac{4(1+\epsilon)}{z}-\frac{2(1+\epsilon)}{z-1}-\frac{2(1+\epsilon)}{z+1}\right)\M_2+
\left(\frac{4\epsilon}{z}-\frac{1+4\epsilon}{z-1}-\frac{1+4\epsilon}{z+1}\right)\M_3 .
\eqa
 Solving the above equations at $\epsilon=0$, we find that the
differential equations for the new basis
\bqa
\GG_{2}&=& m_W^2  \,  \M_2\,, \nonumber\\
\GG_{3}&=& (m_W^2-m_t^2) \,  \M_3-2m_t^2\,  \M_2\,
\eqa
are vanishing at $\epsilon=0$.
Going back to $4-2\epsilon$ dimension, we have
\bqa
\frac{\text{d} \GG_{2} }{\text{d} z}&=&\epsilon\left(\frac{2\GG_{2}}{z}-\frac{2\GG_{2}+\GG_{3}}{z-1}-\frac{2\GG_{2}+\GG_{3}}{z+1}\right),\nonumber\\
\frac{\text{d} \GG_{3} }{\text{d} z}&=&\epsilon\left(\frac{8\GG_{2}}{z}-2\frac{2\GG_{2}+\GG_{3}}{z-1}-2\frac{2\GG_{2}+\GG_{3}}{z+1}\right).
\label{eq:diffeq}
\eqa
where the parameter $\epsilon$ of the spacetime dimension  appears only as a multiplicative factor on the right hand side of the differential equations,
which is called the canonical form or  the $d\ln $ form \cite{Henn:2013pwa}.

In this way,  we obtain the following MIs that obey canonical differential equations.
\bqa
\GG_{1}&=&   m_t^2\M_1\,, \qquad
\GG_{2}= m_W^2  \,  \M_2\,, \qquad
\GG_{3}= (m_W^2-m_t^2) \,  \M_3-2m_t^2\,  \M_2\,,  \nonumber\\
\GG_{4}&=& (-s)\, \M_4\,,  \qquad
\GG_{5}= r_1 \, \M_5 \,,  \qquad
\GG_{6}= (-s)\,\M_6\,,  \qquad
\GG_{7}= r_1 \,  \M_7\,,  \nonumber\\
\GG_{8}&=& m_t^2 r_1 \,\M_8\,,   \qquad
\GG_{9}= m_W^2 s\,  \M_9+m_t^2(m_t^2-m_W^2-s)\,\M_8+\frac{3}{2}(m_t^2-m_W^2-s)\,\M_7\,,  \nonumber\\
\GG_{10}&=&r_1 \,\M_{10}\,,    \qquad
\GG_{11}= m_t^2(-s)\,\M_{11}-\frac{3}{2}(m_t^2-m_W^2+s)\,\M_{10}\,, \nonumber\\
\GG_{12}&=& m_W^2\,s\, \M_{12}\,,    \qquad
\GG_{13}=  s^2\, \M_{13}\,, \qquad
\GG_{14}=(- s)\,r_1 \, \M_{14}\,, \qquad
\GG_{15}= r_1 \,\M_{15}\,,  \nonumber\\
\GG_{16}&=& t \, \M_{16}\,,  \qquad
\GG_{17}= (t-m_t^2)\, \M_{17}-2m_t^2\, \M_{16}\,,  \qquad
\GG_{18}= (m_W^2-s-t) \, \M_{18}\,,  \nonumber\\
\GG_{19}&=& m_t^2(-s) \, \M_{19}\,,\qquad
\GG_{20}= t\,(-s)\M_{20}\,,  \qquad
\GG_{21}= m_t^2(-s)\left((t-m_t^2) \M_{21}-\M_{20}\right)\,,  \nonumber\\
\GG_{22}&=& (t-m_W^2) \, \M_{22}\,, \qquad
\GG_{23}=  (-s) \, \M_{23}\,,  \qquad
\GG_{24}= r_1 \, \M_{24}\,, \nonumber\\
\GG_{25}&=& (t-m_t^2)(-s)\, \M_{25}\,, \qquad
\GG_{26}= (m_t^2-s-t) \, \M_{26}\,,  \nonumber\\
\GG_{27}&=& -(m_W^2\,t-m_t^2(s+t+m_W^2)+m_t^4)\, \M_{27}\,,  \nonumber\\
\GG_{28}&=& (t-m_W^2)(-s) \, \M_{28}\,,  \qquad
\GG_{29}= -(t-m_t^2)s^2\, \M_{29}\,, \qquad
\GG_{30}= (-s)r_1 \, \M_{30}\,,  \nonumber\\
\GG_{31}&=&  s^2 \,( \M_{31}+\M_{14})+s\,\left(-\M_{15}-\M_{10}+2\M_{7}-\frac{3}{2}\M_{5}+3m_t^2\,\M_{8}\right)\nonumber\\
&+&(s+t-m_W^2)\left(s\,\M_{25}-\frac{1}{4}\M_{17}\right)-\frac{s+t-m_W^2}{4(t-m_t^2)}[2(m_t^2+2m_W^2)\,\M_{2}-3s\,\M_{4}\nonumber\\\, &+&(m_t^2-m_W^2)\M_{3}-2(2t+m_t^2) \M_{16}+12(s+t-m_W^2)\M_{18}+8m_t^2\,s\,\M_{19}].
\eqa

The combination coefficients are generally just rational functions in $s,t,m_W^2,m_t^2$,
except the square root product $r_1\equiv \sqrt{s-(m_t-m_W)^2}\sqrt{s-(m_t+m_W)^2}$  in the basis integrals such as $F_5,F_7$.
This  square root  also appears in the differential equations.
It is necessary to first rationalize the square root  before solving the differential equations in terms of multiple polylogarithms.
To achieve this goal, we perform the following change of integration variable,
\beq
s=m_t^2\frac{(x+z)(1+x z)}{x}
\eeq
with $-1<x<1$ so that $r_1 = (1-x)(1+x)z/x$.
Notice that $r_1$ is negative (positive) when $s$ is negative (positive). 
Here we choose $m_W$ or $z$ also as a variable because it is easy to determine the boundary conditions for some integrals at $z=0$.
Then, the differential equations for $\text{{\bf F}}=(\GG_1,\ldots ,\GG_{31})$  can be written as
\bqa
d\, \text{{\bf F}}(x,y,z;\epsilon)=\epsilon\, (d \, \tilde{A})\,  \text{{\bf F}}(x,y,z;\epsilon),
\eqa
with
\bqa
d\, \tilde{A}=\sum_{i=1}^{15} R_i\,  d \ln(\l_i),
\eqa
where $R_i$ are rational matrices.
Their explicit forms are provided  in an auxiliary file.
The arguments $\l_i$ of this $d$\,ln form, which contain all the
dependence of the differential equations on the kinematics, are referred to as the {\it alphabet} and they consist of the following letters,
\begin{align}
\begin{alignedat}{2}
\l_1 & =x\,,&\quad
\l_2 & =x+1\,, \\
\l_3 & =x-1\,, &\quad
\l_4 & =x+z\,, \\
\l_5 & =x\,z+1\,,&\quad
\l_6 & =x~y+z\,,\\
\l_7 & =x\,z+y  \,,&\quad
\l_8 & =y \, , \\
\l_9 & =y-1\,,&\quad
\l_{10} & =y-z^2\, , \\
\l_{11} & =z\,,&\quad
\l_{12} & =z^2-1\, , \\
\l_{13} & =x^2 z+x y+x+z \,,&\quad
\l_{14} & =x^2 z+x \left(y+z^2\right)+z  \, , \\
\l_{15} & =x^2 z+x \left(-y z^2+y+2 z^2\right)+z  \,,&\quad
\l_{16} & =x^2 z+x y+z\,,\\
\l_{17} & =x^2 z^3+x y \left(z^2-1\right)+2 x z^2+z^3 .
\end{alignedat} \stepcounter{equation}\tag{\theequation}
\label{alphabet}
\end{align}
Notice that the last two letters, $\l_{16}$ and $\l_{17}$, only appears for the non-planar integral family discussed in the appendix.

Since the roots of the letters above are purely algebraic,
the solutions of the differential equations can be directly expressed in terms of multiple polylogarithms \cite{Goncharov:1998kja},
which are defined as $G(x)\equiv 1$ and
\bqa
G_{a_1,a_2,\ldots,a_n}(x) &\equiv & \int_0^x \frac{\text{d} t}{t - a_1} G_{a_2,\ldots,a_n}(t)\, ,\\
G_{\overrightarrow{0}_n}(x) & \equiv & \frac{1}{n!}\ln^n x\, .
\eqa
The length $n$ of the vector $(a_1,a_2,\ldots,a_n)$ is referred to as the transcendental $weight$ of multiple polylogarithms.

\section{Boundary conditions and analytical results}
\label{sec:result}

In order to obtain the analytical solutions of the differential equations for the canonical basis shown above, we need to fix the boundary conditions first.

The base $\GG_1$ is obtained by integration directly,
which can also be found in \cite{Chen:2017xqd}.
\beq
 \GG_1=-\frac{1}{4}-\epsilon^2\frac{5\pi^2}{24}-\epsilon^3\frac{11\zeta(3)}{6}-\epsilon^4\frac{101\pi^4}{480}+{\cal O}(\epsilon^{5}).
\label{eq:F2}
\eeq

The loop integrals in the planar family do not have a branch cut at $m_W=0~(z=0)$.
Thus, the corresponding canonical differential equations should not have a pole at $z=0$.
This regularity condition provides useful information about the boundaries.
As can be seen from Eq.(\ref{eq:diffeq}), the coefficient of $1/z$ should vanish at $z=0$,
which means $F_2|_{z=0}=0$.
Due to the same reason, the bases $\GG_9$ and $\GG_{12}$ are also vanishing  at $z=0$,
and
\begin{align}
\GG_{11}|_{z=0}=\left(\GG_{1}-\frac{\GG_{4}}{2}\right)\bigg|_{z=0} \,.
\end{align}
The boundary condition for $\GG_3$ at  $z=0$ is calculated directly,
\begin{align}
\GG_{3}|_{z=0}=1+\epsilon^2\frac{\pi^2}{2}-\epsilon^3\frac{8\zeta(3)}{3}+\epsilon^4\frac{7\pi^4}{40}+{\cal O}(\epsilon^{5}).
\end{align}

In the bases $\{\GG_4,\GG_{23}\}$, the final-state $W$ boson and top quark can be considered as a single particle. All the propagators are massless.
They appear in the massless double box diagrams.
Here we derive independently their values at $s=m_t^2$,
 which can be used as the boundary at $z=0,x=1$.
\begin{align}
\GG_{4}|_{s=m_t^2} & =-1-2\epsilon\, i\, \pi+\epsilon^2\frac{13\pi^2}{6}+\epsilon^3\frac{32\zeta(3)+5 i \pi^3}{3}+\epsilon^4\left(-\frac{101\pi^4}{120}+\frac{64 i\, \pi\, \zeta(3)}{3}\right)+{\cal O}(\epsilon^{5}), \nn \\
\GG_{23}|_{s=m_t^2} & =\frac{1}{4}+\epsilon\, \frac{i\, \pi}{2}-\epsilon^2\frac{11\pi^2}{24}-\epsilon^3\left(\frac{13\zeta(3)}{6}+\frac{ i \pi^3}{4}\right)
+\epsilon^4\left(\frac{79\pi^4}{1440}-\frac{13 i\, \pi\, \zeta(3)}{3}\right)+{\cal O}(\epsilon^{5}) .\nn
\end{align}

The bases $\{\GG_6,\GG_{13}\}$ factorize to a product of two one-loop integrals,
and can be computed easily,
\begin{align}
\GG_{6}|_{s=m_t^2} & =1+\epsilon\, i\, \pi-\epsilon^2\frac{\pi^2}{2}-\epsilon^3\frac{16\zeta(3)+i \pi^3}{3}+\epsilon^4\left(\frac{\pi^4}{120}-\frac{8 i\, \pi\, \zeta(3)}{3}\right)+{\cal O}(\epsilon^{5}), \nn\\
\GG_{13}|_{s=m_t^2} & =1+ 2\epsilon\, i\, \pi-\epsilon^2\frac{13\pi^2}{6}-\epsilon^3\frac{14\zeta(3)+5 i \pi^3}{3}+\epsilon^4\left(\frac{113\pi^4}{120}-\frac{28 i\, \pi\, \zeta(3)}{3}\right)+{\cal O}(\epsilon^{5})\,. \nn
\end{align}

The integrals of $\{\GG_{5},\GG_{7},\GG_{8},\GG_{10},\GG_{14},\GG_{15},\GG_{24},\GG_{30}\}$ are multiplied by $r_1$ in the bases,
and thus they are vanishing at $x=1$.

The bases $\{\GG_{16},\GG_{17}\}$ are the same as $\{\GG_{2},\GG_{3}\}$
after replacing $t$ by $m_W^2$.
So their boundaries at $y=0$ are known from $\{\GG_{2},\GG_{3}\}$ at $z=0$.

From the definitions of the bases, we know that
$\GG_{18},\GG_{22},\GG_{26},\GG_{27}$ vanish at
$u=m_t^2~ (l_{13}=0),t=m_W^2~(y=z^2),u=m_W^2\, (l_{14}=0),m_W^2\,t-m_t^2(s+t+m_W^2)+m_t^4=0~(l_{15}=0)$,
respectively.

The boundary conditions of $\{\GG_{19},\GG_{20},\GG_{21},\GG_{25},\GG_{28},\GG_{29},\GG_{31}\}$ are determined from the regularity conditions at $u\, t=m_t^2\,m_W^2 ~(x=-\frac{y}{z})$.

With the discussion above, we determine all the boundary conditions for the planar family.
As a result, one can obtain the analytic results of the basis from the canonical differential equations directly.
We provide the results of the MIs in electronic form in the ancillary files
attached to the arXiv submission of the paper.
Below we show the first two terms in the expansion of $\epsilon$.
\begin{align}
 \GG_1 & = -\frac{1}{4} + \epsilon \cdot 0 +{\cal O}(\epsilon^2)\,, \qquad
 \GG_2  = 0 - \epsilon \cdot  \ln \left(1-z^2\right)+{\cal O}(\epsilon^2)\,,  \nn\\
 \GG_3 & = 1 - \epsilon \cdot   2 \ln \left(1-z^2\right)+{\cal O}(\epsilon^2)\,,  \nn\\
 \GG_4 & = -1 + \epsilon \cdot   2 \ln \left(\frac{(x+z) (x z+1)}{x}\right)-2 i \pi +{\cal O}(\epsilon^2)\,, \qquad
 \GG_5  = 0 - \epsilon \cdot  0+{\cal O}(\epsilon^2) \,,  \nn\\
 \GG_6 & = 1 - \epsilon \cdot  \ln \left(\frac{(x+z) (x z+1)}{x}\right)+i \pi +{\cal O}(\epsilon^2)\,,  \nn\\
 \GG_7 & = 0 + \epsilon \cdot   0 +{\cal O}(\epsilon^2)\,, \qquad
 \GG_8  = 0 + \epsilon \cdot   0 +{\cal O}(\epsilon^2)\,,  \nn\\
 \GG_9 & = 0 - \epsilon \cdot   \ln \left(1-z^2\right)+{\cal O}(\epsilon^2)\,,  \qquad
 \GG_{10}  = 0 + \epsilon \cdot  0 +{\cal O}(\epsilon^2)\,,  \nn\\
 \GG_{11} & = \frac{1}{4} + \epsilon \cdot  \left[ -\ln \left(\frac{(x+z) (x z+1)}{x}\right)+\ln\left(1-z^2\right)+i \pi \right]+{\cal O}(\epsilon^2)\,,  \nn\\
 \GG_{12}  & = 0 - \epsilon \cdot  \ln \left(1-z^2\right)+{\cal O}(\epsilon^2)\,,  \nn\\
 \GG_{13} & = 1 + \epsilon \cdot  \left[ -2 \ln \left(\frac{(x+z) (x z+1)}{x}\right)+2 i \pi  \right]+{\cal O}(\epsilon^2)\,,  \qquad
 \GG_{14}  = 0 + \epsilon \cdot  0+{\cal O}(\epsilon^2)\,,  \nn\\
 \GG_{15} & = 0 + \epsilon \cdot   0 +{\cal O}(\epsilon^2)\,,  \qquad
 \GG_{16}  = 0 - \epsilon \cdot    \ln (1-y)+{\cal O}(\epsilon^2)\,,  \nn\\
 \GG_{17} & = 1 - \epsilon \cdot  2 \ln(1-y)+{\cal O}(\epsilon^2)\,,  \qquad
 \GG_{18}  = 0 + \epsilon \cdot   0+{\cal O}(\epsilon^2)\,,  \nn\\
 \GG_{19} & = -\frac{1}{6} + \epsilon \cdot \left[ \frac{1}{2} \ln \left(\frac{(x+z) (x z+1)}{x}\right)-\frac{1}{3} \ln (1-y)-\frac{i \pi }{2}\right]+{\cal O}(\epsilon^2)\,,  \nn\\
 \GG_{20}  & = 0 - \epsilon \cdot   \ln (1-y)+{\cal O}(\epsilon^2)\,,  \nn\\
 \GG_{21} & = \frac{5}{8} + \epsilon \cdot   \left[ -\frac{1}{2} \ln \left(\frac{(x+z) (x z+1)}{x}\right)-\ln (1-y)+\frac{1}{2} \ln \left(1-z^2\right)+\frac{i \pi }{2} \right]+{\cal O}(\epsilon^2)\,,  \nn\\
 \GG_{22} & = 0 + \epsilon \cdot     \left[  \frac{1}{2} \ln (1-y)-\frac{1}{2} \ln \left(1-z^2\right) \right]+{\cal O}(\epsilon^2)\,,  \nn\\
 \GG_{23} & = \frac{1}{4} + \epsilon \cdot    \left[  -\frac{1}{2} \ln \left(\frac{(x+z) (x z+1)}{x}\right)+\frac{i \pi }{2}\right]+{\cal O}(\epsilon^2)\,,  \qquad
 \GG_{24}  = 0 + \epsilon \cdot   0+{\cal O}(\epsilon^2)\,,  \nn\\
 \GG_{25} & = \frac{5}{12} + \epsilon \cdot    \left[ -\frac{1}{2} \ln
   \left(\frac{(x+z) (x z+1)}{x}\right)-\frac{7}{6} \ln (1-y)+\frac{1}{2} \ln \left(1-z^2\right)+\frac{i \pi }{2} \right]+{\cal O}(\epsilon^2)\,,  \nn \\
 \GG_{26}  & = 0 + \epsilon \cdot   0 +{\cal O}(\epsilon^2)\,,  \qquad
 \GG_{27}   = 0 +  \epsilon \cdot  0 +{\cal O}(\epsilon^2)\,,  \nn \\
 \GG_{28}  & = 0 + \epsilon \cdot    \left[ \frac{1}{2} \ln(1-y)-\frac{1}{2} \ln \left(1-z^2\right) \right]+{\cal O}(\epsilon^2)\,,  \nn\\
 \GG_{29} & = -\frac{11}{24} + \epsilon \cdot   \left[ \frac{1}{2} \ln \left(\frac{(x+z) (x z+1)}{x}\right)+\frac{4}{3} \ln (1-y)-\frac{1}{2}
   \ln \left(1-z^2\right)-\frac{i \pi }{2} \right]+{\cal O}(\epsilon^2)\,,  \nn \\
 \GG_{30} & = 0 + \epsilon \cdot  0 +{\cal O}(\epsilon^2)\,,  \qquad
 \GG_{31}  = \frac{1}{24} - \epsilon \cdot  \frac{1}{6} \ln (1-y)+{\cal O}(\epsilon^2)\,.
 \label{eq:anares}
\end{align}

In our calculation, we have varied $m_W$ in order to choose a proper boundary condition.
One may wonder whether  the boundary $m_W=m_t$, equivalently $z=1$, can be taken.
If the answer is yes, then one can make use of all the results of two-loop integrals for $gg\to t\bar{t}$.
However, this is non-trivial since $z=1$ is the point where a branch cut starts.
For example, we can take $F_1$ as a boundary for $F_2$ at $z=1$ because  they are the same if setting $m_W^2=m_t^2$ in the integrands.
But we see from above analytic results expanded in $\epsilon$ that ${F_2}|_{z=1}\neq F_1$.
The reason is that the analytic results are valid  only for $z^2<1$.
In the $z\to1$ limit, one can not expand the $(1-z)^{n\epsilon}$ terms in a series of $\epsilon$ for the master integrals. 
Instead, one should now solve the differential equation in Eq.(\ref{eq:diffeq}) with full $\epsilon$ dependence,
\begin{align}
 \GG_{2}  & = c_1 (1-z)^{-4\epsilon} -c_2,  \nn \\
  \GG_{3}  & = 2c_1 (1-z)^{-4\epsilon} +2c_2  .
\end{align}
Comparing these with the analytic results in Eq.(\ref{eq:anares}), we find
\begin{align}
c_1= \frac{1}{4}-\epsilon \ln 2+{\cal O}(\epsilon^2) ,\quad c_2 =\frac{1}{4}+{\cal O}(\epsilon^2) .
\end{align}
Then taking $(1-z)^{-4\epsilon}\to 0$ at $z=1$, we see that ${ \GG_2}|_{z=1}=  \GG_1$.
If one uses the boundary values at $z=1$ for $\GG_2$ and $\GG_3$, which are $-c_2$ and $2c_2$, respectively,
one still needs the information of $c_1$ to obtain the results at general  $z$.
But this information can only be obtained at a point other than $z=1$.

All the analytic results are real in the Euclidean  regions $(s<0,t<0,u<0)$.
 In this work we are interested in the physical region with $s>(m_t+m_W)^2,t_0<t<t_1,0<m_W^2<m_t^2$, 
 \footnote{Notice that the physical region with $0<s<(m_t-m_W)^2,t_1<t<t_0,0<m_W^2<m_t^2$ corresponds to top quark decay $t\to Wbg$,
and our prescription for the analytic continuation is applicable in this case. }
where
 \begin{align}
 t_0&\equiv \frac{m_t^2+m_W^2-s-r_1}{2}, \nn\\
 t_1&\equiv \frac{m_t^2+m_W^2-s+r_1}{2}.
 \end{align}
This region corresponds to $0<x<1, -2z/x<y<-2zx, 0<z<1$.
The analytic continuation to this region can be performed by assigning $s$  a numerically small imaginary part $i\varepsilon\,(\varepsilon>0)$, i.e., $s\rightarrow s+i\varepsilon$.
This prescription gives the correct numerical results in both  the Euclidean and physical regions, when the multiple polylogarithms are evaluated using {\tt GiNaC} \cite{Vollinga:2004sn,Bauer:2000cp}.

All the analytical results have been checked  with  the numerical package {\tt FIESTA} \cite{Smirnov:2015mct},
and they agree  within the computation errors in both Euclidean and physical regions.
For example, we show the results of two integrals at a physical kinematic point $(s=10,t=-2,m_W^2=\frac{1}{4},m_t=1)$,
\bqa
I_{1, 0, 1, 0, 1, 1, 1, 0, 0}^{\rm analytic}&=&\frac{0.00475421+ 1.48022009\,i}{\epsilon}\nonumber\\
&+&(-5.24410651+1.22399295\,i),\\
I_{1, 0, 1, 0, 1, 1, 1, 0, 0}^{\rm FIESTA}&=&\frac{0.004754+1.48022\, i\pm0.000056(1+i) }{\epsilon}\nonumber\\
&+&(-5.24410 + 1.22399 \,i) \, \pm (0.000416+0.000415\,i)\,,
\eqa
and
 \bqa
I_{1, 1, 1, 1, 1, 0, 1, 0, 0}^{\rm analytic}&=&\frac{0.0308065}{\epsilon^3}+\frac{-0.06040731}{\epsilon^2}+\frac{0.22341495-0.06475586\,i}{\epsilon}\nonumber\\&+&(-0.26302494 +0.62749975\,i),\\
I_{1, 1, 1, 1, 1, 0, 1, 0, 0}^{\rm FIESTA}&=&\frac{0.030807  \pm0.000005 }{\epsilon^3}+\frac{-0.060407  \pm0.000027 }{\epsilon^2}\nonumber\\
&+&\frac{0.223415 -0.064756\, i  \pm(0.000116+0.000124 \,i) }{\epsilon}\nonumber\\
&+&(-0.263019 + 0.627484 \,i) \, \pm (0.000392 + 0.000395 \,i).
\eqa

\section{Conclusion}
\label{sec:conclusion}
We calculate analytically two-loop master integrals for hadronic $tW$  productions that contain only one massive propagator.
After choosing a canonical basis, the differential equations for the master integrals can be transformed into the $d$\,ln form.
The boundaries  are determined by simple direct integrations or
regularity conditions at kinematic points without physical singularities.
The analytical results in this work are expressed in terms of multiple polylogarithms,
and have been checked with numerical computations.
There is still a lot of work to do in the future to obtain
the complete two-loop  virtual corrections in this channel.

\section*{Acknowledgments}

This work was supported in part by the National Natural Science Foundation of China under Grant No. 11805042, 12005117, and 12175048.
The work of JW was also supported 
by Taishan Scholar Foundation of Shandong province (No. tsqn201909011).

\section*{Appendix: Results of the non-planar integral family}
\label{sec:app}

For the master integral shown in Fig.\ref{fig:massive1}$(b)$,
we define the non-planar integral family  by
\bqa
J_{n_1,n_2,\ldots,n_{9}}=\int{\mathcal D}^D q_1~{\mathcal D}^D q_2\frac{1}{P_1^{n_1}~P_2^{n_2}~P_3^{n_3}~P_4^{n_4}~P_5^{n_5}~P_6^{n_6}~P_7^{n_7}P_8^{n_8}~P_9^{n_9}}
\label{def}
\eqa
with the denominators
\bqa
P_1&=& q_1^2,\quad P_2=(q_1-q_2)^2,\quad P_3=q_2^2,\quad P_4=(q_1+k_1)^2,\nonumber\\
P_5& =& (q_1-q_2-k_2)^2,\quad P_6=(q_2+k_1+k_2)^2,\quad P_7=(q_2+k_1+k_2-k_3)^2-m_t^2,\nonumber\\
P_8&=&(q_1-k_3)^2,\quad P_9=(q_2+k_1)^2.\nonumber
\label{int2}
\eqa

The canonical bases are chosen to be
\bqa
\BB_{1}&=&   m_t^2\N_1\,,\quad \BB_{2}= m_W^2  \,  \N_2\,,\quad \BB_{3}= (m_W^2-m_t^2) \,  \N_3-2m_t^2\,  \N_2\,, \nonumber\\
\BB_{4}&=& (-s)\, \N_4\,, \quad \BB_{5}= r_1 \, \N_5 \,, \quad \BB_{6}= (m_W^2+m_t^2-s-t)\,\N_6\,,   \nonumber\\
\BB_{7}&=& (m_W^2-s-t)\,\N_7-2m_t^2\,\N_6\,,  \quad \BB_{8}= (m_W^2-s-t)\,\N_8\,, \nonumber\\
\BB_{9}&=&  s\,  \N_9\,,\quad \BB_{10}= t\,\N_{10}\,,  \quad 
\BB_{11}= (t-m_t^2)\,\N_{11}-2m_t^2\,\N_{10}\,, \nonumber\\
\BB_{12}&=& (t-m_W^2)\, \N_{12}\,,  \quad
\BB_{13}= r_1 \, \N_{13}\,, \nonumber\\
\BB_{14}&=& m_t^2(-s)\, \N_{14}-\frac{3}{2}(m_t^2-m_W^2+s)\,\N_{13}\,, \nonumber\\
\BB_{15}&=& r_1  \,\N_{15}\,,  \quad
\BB_{16}= s(s+t-m_W^2) \, \N_{16}\,,  \quad
\BB_{17}= (t-m_t^2)\, \N_{17}\,,  \nonumber\\
\BB_{18}&=& m_t^2(-s) \, \N_{18}\,,  \quad
\BB_{19}= r_1  \, \N_{19}\,,  \quad
\BB_{20}= (t-m_t^2)(-s) \, \N_{20}\,,  \nonumber\\
\BB_{21}&=& (m_W^2-s-t)\,\N_{21}\,,   \quad
\BB_{22}= m_t^2(-s)\,\N_{22}\,,   \quad
\BB_{23}= (m_t^2-s-t)\,\N_{23}\,,   \nonumber\\
\BB_{24}&=& -(t\, m_W^2-(m_W^2+s+t)m_t^2+m_t^4)\,\N_{24}\,,   \quad
\BB_{25}= (t-m_W^2)\,\N_{25}\,,   \nonumber\\
\BB_{26}&=& (m_W^2(s+t-m_W^2)-m_t^2(t-m_W^2))\,\N_{26}\,,   \quad
\BB_{27}= (-s)\,\N_{27}\,,   \nonumber\\
\BB_{28}&=& (t-m_t^2)(m_W^2-s-t)\,\N_{28}\,,  \quad
\BB_{29}= (m_W^2-m_t^2)s\,\N_{29}\,,   \nonumber\\
\BB_{30}&=& (t-m_W^2)\,\N_{30}+(m_W^2-s-t)\,\N_{27}\,,   \quad
\BB_{31}= s^2\,\N_{31}\,,   \nonumber\\
\BB_{32}&=& (s+t-m_W^2)\left(s^2\,\N_{32}+s\,\N_{33}-s\,\N_{29}+\frac{1}{4}(s+t-m_t^2)\N_{28}+\frac{\N_{11}}{8}\right)\nonumber\\
&+&\frac{(s+t-m_W^2)}{(t-m_t^2)}\bigg(\frac{3}{2} \N_{21} \left(-m_W^2+s+t\right)+\N_{22} s m_t^2+\frac{1}{4} \N_2 \left(m_t^2+2 m_W^2\right)\nonumber\\
&+&\frac{1}{8} \N_3 \left(m_t^2-m_W^2\right)-\frac{1}{4} \N_{10} \left(m_t^2+2 t\right)-\frac{3 \N_4 s}{8}\bigg)\nonumber\\
&+&\frac{1}{4\epsilon+1}\bigg[-\frac{1}{8} \left(2 \N_{28} s+\N_7+\N_{11}\right) \left(-m_W^2+s+t\right)+\N_{18} s m_t^2\nonumber\\
&+&\frac{1}{4} \N_6 \left[2 \left(-m_W^2+s+t\right)-3 m_t^2\right]+\frac{3}{2} \N_{17} \left(m_t^2-t\right)\nonumber\\
&+&\frac{s+t-m_W^2}{t-m_t^2}\left(-\frac{3}{2} \N_{21} \left(-m_W^2+s+t\right)+\N_{22} (-s) m_t^2+\frac{1}{4} \N_{10} \left(m_t^2+2 t\right)\right)\nonumber\\
&+&\frac{s+m_t^2-m_W^2}{t-m_t^2}\left(-\frac{1}{4} \N_2 \left(m_t^2+2 m_W^2\right)+\frac{1}{8} \N_3 \left(m_W^2-m_t^2\right)+\frac{3 \N_4 s}{8}\right)\bigg]\nonumber\\
\BB_{33}&=& (t-m_t^2)(-s)\,\N_{33}\,,   \nonumber\\
\BB_{34}&=& r_1 \,\bigg[\N_{34}+s\,\N_{33}-\N_{30}
-\frac{1}{4}(s+t-m_W^2)\N_{28}+\frac{1}{2}\N_{17}-\frac{1}{12}\N_{11}\nonumber\\
&+&\frac{1}{t-m_t^2}\bigg(\frac{m_t^2}{4}\N_{1}-\frac{m_t^2+2m_W^2}{4}\N_{2}-\frac{m_t^2-m_W^2}{8}\N_{3}\nonumber\\
&+&\frac{3\, s}{8}\N_{4}+\frac{2t+m_t^2}{6}\N_{10}-\frac{3}{2}(s+t-m_W^2)\N_{21}-m_t^2\,s\N_{22}\bigg)\bigg]\,.
\eqa
with
\begin{align*}
\N_{1}&=\epsilon^2 \, J_{1, 2, 0, 0, 0, 0, 2, 0, 0}\,,  &
\N_{2}&=\epsilon^2 \, J_{0, 0, 0, 1, 2, 0, 2, 0, 0}\,,  &
\N_{3}&=\epsilon^2 \, J_{0, 0, 0, 2, 2, 0, 1, 0, 0}\,,  \\
\N_{4}&=\epsilon^2 \, J_{0, 0, 1, 2, 2, 0, 0, 0, 0}\,,  &
\N_{5}&=\epsilon^3 \, J_{0, 0, 1, 1, 2, 0, 1, 0, 0}\,,  &
\N_{6}&=\epsilon^2 \, J_{0, 1, 0, 2, 0, 0, 2, 0, 0}\,,  \\
\N_{7}&=\epsilon^2 \, J_{0, 2, 0, 2, 0, 0, 1, 0, 0}\,,  &
\N_{8}&=\epsilon^3 \, J_{0, 1, 0, 2, 0, 1, 1, 0, 0}\,,  &
\N_{9}&=\epsilon^3 \, J_{0, 1, 1, 2, 0, 1, 0, 0, 0}\,,  \\
\N_{10}&=\epsilon^2 \, J_{1, 0, 0, 0, 2, 0, 2, 0, 0}\,,  &
\N_{11}&=\epsilon^2 \, J_{2, 0, 0, 0, 2, 0, 1, 0, 0}\,,  &
\N_{12}&=\epsilon^3 \, J_{1, 0, 0, 0, 2, 1, 1, 0, 0}\,,  \\
\N_{13}&=\epsilon^3 \, J_{1, 2, 0, 0, 0, 1, 1, 0, 0}\,,  &
\N_{14}&=\epsilon^2 \, J_{1, 2, 0, 0, 0, 1, 2, 0, 0}\,,  &
\N_{15}&=\epsilon^3(1-2\epsilon) \, J_{0, 1, 1, 1, 0, 1, 1, 0, 0}\,,  \\
\N_{16}&=\epsilon^3 \, J_{0, 1, 1, 2, 0, 1, 1, 0, 0}\,,  &
\N_{17}&=\epsilon^4 \, J_{0, 1, 1, 1, 1, 0, 1, 0, 0}\,,  &
\N_{18}&=\epsilon^3\, J_{0, 1, 1, 1, 1, 0, 2, 0, 0}\,,  \\
\N_{19}&=\epsilon^3(1-2\epsilon)\, J_{1, 0, 1, 0, 1, 1, 1, 0, 0}\,,  &
\N_{20}&=\epsilon^3 \, J_{1, 0, 1, 0, 2, 1, 1, 0, 0}\,, &
\N_{21}&=\epsilon^4 \, J_{1, 0, 1, 1, 1, 0, 1, 0, 0}\, , \\
\N_{22}&=\epsilon^3\, J_{1, 0, 1, 1, 1, 0, 2, 0, 0}\,,  &
\N_{23}&=\epsilon^4 \, J_{1, 1, 0, 0, 1, 1, 1, 0, 0}\,, &
\N_{24}&=\epsilon^3\, J_{1, 1, 0, 0, 1, 1, 2, 0, 0} \, , \\
\N_{25}&=\epsilon^4\, J_{1, 1, 0, 1, 0, 1, 1, 0, 0}\,,  &
\N_{26}&=\epsilon^3 \, J_{1, 1, 0, 1, 0, 1, 2, 0, 0}\,, &
\N_{27}&=\epsilon^4 \, J_{1, 1, 0, 1, 1, 0, 1, 0, 0}\, , \\
\N_{28}&=\epsilon^3\, J_{1, 1, 0, 1, 1, 0, 2, 0, 0}\,,  &
\N_{29}&=\epsilon^4 \, J_{1, 1, 0, 1, 1, 1, 1, 0, 0}\,, &
\N_{30}&=\epsilon^4 \, J_{1, 1, 0, 1, 1, 1, 1, 0, -1}\, , \\
\N_{31}&=\epsilon^4\, J_{1, 1, 1, 1, 1, 1, 0, 0, 0}\,, &
\N_{32}&=\epsilon^4\, J_{1, 1, 1, 1, 1, 1, 1, 0, 0}\,,\\
\N_{33}&=\epsilon^4\, J_{1, 1, 1, 1, 1, 1, 0, 0, -1}\,, &
\N_{34}&=\epsilon^4\, J_{1, 1, 1, 1, 1, 1, 1, 0, -2}\,.
\stepcounter{equation}\tag{\theequation}
\label{def:LBasisT2}
\end{align*}

The canonical differential equations for $\text{{\bf B}}=(\BB_1,\ldots, \BB_{34})$  can be written as
\bqa
d\, \text{{\bf B}}(x,y,z;\epsilon)=\epsilon\, (d \, \tilde{C})\,  \text{{\bf B}}(x,y,z;\epsilon),
\eqa
with
\bqa
d\, \tilde{C}=\sum_{i=1}^{17} Q_i\,  d \ln(\l_i),
\eqa
where $Q_i$ are rational matrices.

The non-planar and planar diagrams share some common integrals.
For the non-planar family, we find that
\begin{align}
\BB_1 & =\GG_1\, ,\quad
\BB_2  =\GG_2\, ,\quad
\BB_3  =\GG_3\, , \quad
\BB_4  =\GG_4\, ,\nn \\
\BB_5 &=\GG_5\, ,\quad
\BB_9 =-\GG_{23}\, ,\quad
\BB_{10} =\GG_{16}\, ,\quad
\BB_{11} =\GG_{17}\, ,\nn \\
\BB_{12} &=\GG_{22}\, ,\quad
\BB_{13} =\GG_{10}\, ,\quad
\BB_{14} =\GG_{11}\, ,\quad
\BB_{19} =\GG_{24}\, ,\nn \\
\BB_{20} &=\GG_{25}\, ,\quad
\BB_{23} =\GG_{26}\, ,\quad
\BB_{24} =\GG_{27}\, .
\label{brelation}
\end{align}

For the other unknown integrals in the non-planar family, their boundary conditions are obtained as follows.
The base $\BB_{6}$ is vanishing at $u=0\, (l_{16}=0)$,
and the boundary conditions for $\BB_{7}$ at $u=0$ are equal to $\BB_{3}$ at $m_W=0$. The base $\BB_{8}$ vanishes at $u=m_W^2$. 
The bases $\{ \BB_{15},\BB_{34}\}$ vanish at $s=(m_t+m_W)^2$.
The base $\BB_{17}$ is vanishing at $t=m_t^2$. 
The base $\BB_{21}$ equals to zero at $u=m_t^2$.
 The  base $\BB_{27}$ is vanishing at $s=0$.
 The base $\BB_{30}$ is zero at $u=m_W^2$. 
 The base $\BB_{26}$ equals to zero at $m_W^2(s+t-m_W^2)-m_t^2(t-m_W^2)=0$, i.e. $l_{17}=0$. 
 The result of $\BB_{31}$ can be found  in Ref.\cite{Gonsalves:1983nq}.
 The boundary conditions for bases $\{ \BB_{16},\BB_{18},\BB_{22},\BB_{25},\BB_{28},\BB_{29},\BB_{32},\BB_{33}\}$ are determined from the regularity conditions at $u\, t=m_t^2\, m_W^2$.
The analytical results are expressed in terms of multiple polylogarithms.
We provide them in the ancillary file, which can be evaluated using {\tt GiNaC}.
In the physical region, one needs to assign $s$ and $t$  
a numerically small but positive imaginary part.

\bibliography{tw-paper}
\bibliographystyle{JHEP}

\end{document}